\documentclass{article}

\usepackage[preprint,nonatbib]{neurips_2025}

\usepackage{cite}

\usepackage[utf8]{inputenc} 
\usepackage[T1]{fontenc}    
\usepackage{hyperref}       
\usepackage{url}            
\usepackage{booktabs}       
\usepackage{amsfonts}       
\usepackage{nicefrac}       
\usepackage{microtype}      
\usepackage{xcolor}         
\usepackage{enumitem}
\usepackage[ruled,vlined]{algorithm2e}
\usepackage{amsmath}
\usepackage{amssymb}
\usepackage{wrapfig} 
\usepackage{graphicx}
\usepackage{multirow}
\usepackage{pifont}
\usepackage{multicol}
\usepackage[table]{xcolor}
\usepackage{colortbl}
\usepackage{caption}

\title{CoreUnlearn: Rethinking Concept Unlearning through Disentangled Component-Level Erasure in Text-guided Diffusion Models}

\author{Mengnan Zhao, Lihe Zhang, Baocai Yin} 

\begin{document}

\maketitle

\begin{abstract}
Text-guided diffusion models have revolutionized image synthesis but also raise ethical concerns, such as privacy violation and harmful content generation. To mitigate these issues, prevailing methods typically leverage an alignment mechanism—with predefined erasure references—to fine-tune pretrained model weights.
However, these techniques are intrinsically limited by the representational capacity of textual space and display high sensitivity to the choice of predefined erasure references, $e.g.$, suboptimal references may significantly affect the model utility preservation during erasure.
To overcome these limitations, we introduce CoreUnlearn, aiming to disentangle and remove the erasure-critical component of the undesirable concept.
Specifically, CoreUnlearn comprises a Component Extraction Module (CEM) and a Swap Disentangling Strategy (SDS). Guided by SDS, CEM is pre-trained to decompose concept embeddings into distinct component types. Leveraging this decomposition, CoreUnlearn then removes the erasure-critical component while retaining non-critical ones by fine-tuning model weights. Extensive experiments demonstrate that CoreUnlearn achieves effective concept erasure with minimal impact on overall model performance.

\end{abstract}

\section{Introduction}
Text-to-image generative techniques \cite{saharia2022photorealistic,zhang2023adding,ruiz2023dreambooth}, especially text-guided diffusion models \cite{xing2024svgdreamer,meng2023distillation}, have transformed image synthesis by producing high-quality outputs across diverse domains \cite{kim2025race,lyu2024one}, such as digital art, media content, and healthcare \cite{croitoru2023diffusion,yang2023diffusion}.  
However, despite their remarkable success, these models often inherit biases or sensitive information from pre-training datasets, raising ethical concerns such as copyright infringement, privacy breaches, and the generation of harmful content \cite{zhao2024separable,lu2024mace,gong2025reliable,jain2024trasce}. 
To mitigate these issues, retraining-based methods like Safe Latent Diffusion \cite{schramowski2023safe} have been proposed.

Unlike retraining-based approaches, machine unlearning (MU) \cite{gao2024meta, sun2024attentive, hao2025conceptexpress} leverages targeted or non-targeted fine-tuning to suppress the generation of undesirable content in diffusion models.
Targeted fine-tuning \cite{hong2024all} steers the model predictions for undesirable concepts toward predefined textual or semantic anchors, while non-targeted fine-tuning \cite{zhang2024forget,zhao2024separable} directly suppresses the features associated with undesirable concepts.
The predefined anchors in targeted fine-tuning techniques range from empty prompts \cite{kim2023towards} and broad conceptual categories \cite{kumari2023ablating} to partial embedding values derived from text prompts \cite{brack2023sega,hong2024all} or weighted combinations of multiple text embeddings \cite{gandikota2024unified}.

Despite recent advancements in unlearning techniques for diffusion models \cite{seo2024generative,wu2024unlearning,fuchi2024erasing}, the preservation performance of model utility during erasure can be seriously compromised by issues like suboptimal target anchors or unnecessary feature removal
\cite{park2024direct,moon2024holistic,chavhan2024conceptprune,sridhar2024prompt}.
Inspired by this, we consider one question: \textit{Can we disentangle and remove erasure-critical components of undesirable concepts?}  
By removing such components while retaining others, we aim to mitigate the adverse effects of concept erasure on the overall model performance.

To achieve this, we present CoreUnlearn, integrating the Component Extraction Module (CEM) and the Swap Disentangling Strategy (SDS).
Specifically, we first formalize the paradigm for identifying and removing erasure-critical components. We then architect the CEM and define its SDS-based supervision objective.
Guided by SDS, CEM is pre-trained to disentangle embeddings of undesirable concepts into distinct component types.
Finally, we fine-tune the model weights to remove the erasure-critical components selected by pre-trained CEM while preserving all remaining information.

The contributions are four-fold:  
(1) To mitigate the decline of overall model performance during erasure, we introduce the paradigm of erasure-critical component removal.  
(2) We propose the Component Extraction Module to decompose concept embeddings into distinct components.
(3) To identify the erasure-critical component during the decomposition of CEM, we design the Swap Disentangling Strategy.
(4) Extensive experiments on multiple benchmarks demonstrate that CoreUnlearn effectively preserves overall model performance while realizing effective concept erasure.

\section{Proposed Method}
\label{gen_inst}
\subsection{Preliminaries}
Given the concept embeddings $e$, the visual noise \(z \sim \mathcal{N}(0,\mathbf{I})\), let \( f_t(e, z; \theta):\mathcal{E} \times \mathcal{Z} \to \mathcal{X} \) denote the model prediction at timestep \( t \in \{1, \dots, T\} \). The unlearning technique aims to erase undesirable concepts \( c_u \) (embeddings \( e_u \)) while preserving generation quality for retained concepts \( c_n\in \mathcal{C}_{\text{RCs}} \) (embeddings \( e_n\in \mathcal{E}_{\text{RCs}} \)), expressed as
\begin{equation}\label{eq:recent}
\min_{\theta_{\text{op}}} \mathbb{E}_{z \in \mathcal{Z}_D} \left[\left\| f_t(e_u, z; \theta_{\text{op}}) - f_t(e_p, z; \theta_0) \right\|_2 + \sum_{e_n \sim \mathcal{E}_{\text{RCs}}} \| f_t(e_n, z; \theta_{\text{op}}) - f_t(e_n, z; \theta_0)  \|_2 \right],
\end{equation}
where \( \theta_0 \) represents the pre-trained model weights.
\( \theta_{\text{op}}\) denotes optimizable parameters, initialized as \(\theta_{\text{0}}\). 
\(\mathcal{Z}_D\) is the visual dataset.
$e_p$ means the predefined target anchors for erasing the concept $c_u$.
1) The left term of Eq. \ref{eq:recent} alters the model generations for \(c_u\). Meanwhile, its initial value exhibits an inverse correlation with model utility preservation \cite{zhao2024advanchor}. 
2) The right term aims to preserve the overall model performance, while researchers should carefully select the retained concepts $\mathcal{C}_{\text{RCs}}$.
These observations motivate an issue: \textit{Can we construct a target candidate \(f_{\mathrm{cand}}^t\) that simultaneously realizes effective concept erasure and preserves model utility (Shallower Erase),} such that
   \begin{equation}\label{eq:c1}
   \left\| f_t(e_u, z; \theta_0) - f_{\text{cand}}^t \right\|_2 \ll \left\| f_t(e_u, z; \theta_0) - f_t(e_p, z; \theta_0) \right\|_2, \, and \, \arg\max \, g(f_{\text{cand}}^t) \neq \ell_u,
   \end{equation}
where \(\arg\max g(f_{\mathrm{cand}}^t)\) yields the label index predicted by the fixed classifier \(g(\cdot)\) when applied to candidate features, and \(\ell_u\) denotes the ground-truth label index of \(c_u\).


\subsection{Paradigm of Erasure-critical Component Removal}
We employ the latent difference \( \Delta f_{u}^{t,\theta} \triangleq f_t(e_u,z;\theta) - f_t(e_\emptyset,z;\theta) \) to represent the arbitrary undesirable concept $c_u$, \textit{i.e.}, the generation differences introduced by the guidance of $c_u$.
$e_\emptyset$ means the embeddings of the empty prompt.

To construct the candidate $f_{cand}^t$ that satisfies Eq. \ref{eq:c1}, this paradigm first decomposes \( \Delta f_{u}^{t,\theta}\) into distinct components, and then replaces the erasure-critical component.
Specifically, let \( \mathcal{T}: \mathcal{X} \to \mathcal{X}^M \) be a decomposition operator that partitions the latent difference \(\Delta f_{u}^{t,\theta} \) into \( M \) distinct components:
\begin{equation}\label{eq:decomp}
\Delta f_{u}^{t,\theta} = \sum_{j=1}^M \mathcal{T}(\Delta f_{u}^{t,\theta})_j.
\end{equation}
Let \( k \in \{0,\dots,M\} \) index the erasure-critical component, \(f_{\text{cand}}^t\) is further formulated as follows,
\begin{equation}\label{eq:approx}
f_{\text{cand}}^t = f_t(e_u,z;\theta_0) - \mathcal{T}(\Delta f_{u}^{t,\theta_0})_k + \mathcal{T}^\text{new}_k.
\end{equation}
Eq. \ref{eq:approx} reveals that the critical component \( \mathcal{T}(\Delta f_{u}^{t,\theta_0})_k \) for representing \( c_u \) is replaced by a new component \( \mathcal{T}^\text{new}_k \).
Combined with Eq. \ref{eq:approx} and Eq. \ref{eq:c1}, the decomposition operator $\mathcal{T}$ and the optimizable model parameters $\theta_{op}$ should satisfy the following conditions:
   \begin{equation}\label{eq:candidate}
   \left\| \mathcal{T}^\text{new}_k - \mathcal{T}(\Delta f_{u}^{t,\theta_0})_k \right\|_2 \ll \left\| f_t(e_u, z; \theta_0) - f_t(e_p, z; \theta_0) \right\|_2, \, and \,\arg\max \, g(f_{\text{cand}}^t) \neq \ell_u.
   \end{equation}
Typically, the target anchors $e_p$ are closer to $e_u$ than the empty prompt embeddings $e_\emptyset$, that is,
\begin{equation}
    \left\| f_t(e_u, z; \theta_0) - f_t(e_p, z; \theta_0) \right\|_2 \leq \left\| f_t(e_u, z; \theta_0) - f_t(e_\emptyset, z; \theta_0) \right\|_2 = \|\Delta f_{u}^{t,\theta_0}\|_2.
\end{equation}
Hence, we represent the final conditions as follows:
   \begin{equation}\label{eq:candidate2}
   \left\| \mathcal{T}^\text{new}_k - \mathcal{T}(\Delta f_{u}^{t,\theta_0})_k \right\|_2 \ll \|\Delta f_{u}^{t,\theta_0}\|_2,\, and \,\arg\max \, g(f_{\text{cand}}^t) \neq \ell_u.
   \end{equation}


 \subsection{Learning Critical Component Extraction}\label{sec2.3}

In this subsection, we describe how to construct the decomposition operator $\mathcal{T}(\cdot)$. We propose a Component Extraction Module (CEM) and a Swap Disentangling Strategy (SDS). CEM decomposes each concept representation into distinct components, while SDS enforces that each component exhibits a specific type. The pretrained CEM is utilized as $\mathcal{T}(\cdot)$.

The encoder in CEM (CEM-E) compresses the input (\(\Delta f_{u}^{t,\theta_0}\)) into several groups of feature vectors, while the decoder (CEM-D) then reconstructs the input from these vectors, expressed as
\[
    f_{\text{out}} = \text{CEM-D}\left( \text{CEM-E}\left(\Delta f_{u}^{t,\theta_0}\right) \right).
\]
\begin{wrapfigure}{r}{0.5\textwidth}
  \centering
  \vspace{-2mm}
  \includegraphics[width=0.5\textwidth]{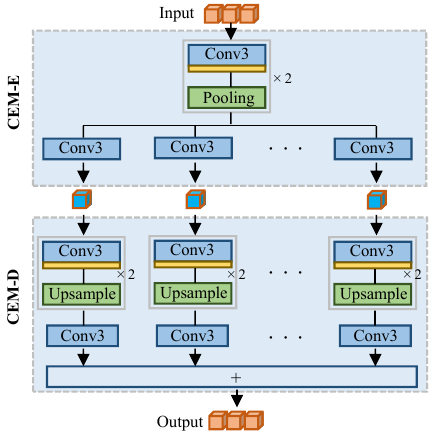}
  \caption{Component extraction module (CEM).}
  \label{PFE}
  \vspace{-8mm}
\end{wrapfigure}
The structure of CEM is illustrated in Figure \ref{PFE}. CEM-D reconstructs several kinds of components, which are summed to produce \( f_\text{out} \). \( f_\text{out} \) is expected to approximate the input \( \Delta f_{u}^{t,\theta_0} \).

Although CEM effectively decomposes the input, it cannot explicitly specify the type of each decomposed component. This limitation raises an issue:  
\textit{
How can we specify the component types and identify which type is most strongly associated with the undesirable concept \(c_u\)?}

Empirically, in text-guided diffusion models, features strongly related to $c_u$ should be robust to visual variations while being sensitive to textual changes. 
Guided by this assumption, we divide component types into four categories:  
\begin{itemize}[leftmargin=10pt, itemsep=-2pt, topsep=0pt]
    \item[-] \textit{$V_rT_r$}: Visual-robust and text-robust,  
    \item[-] \textit{$V_rT_s$}: Visual-robust and text-sensitive,  
    \item[-] \textit{$V_sT_r$}: Visual-sensitive but text-robust,  
    \item[-] \textit{$V_sT_s$}: Visual-sensitive and text-sensitive.  
\end{itemize}

\begin{figure*}[t]
\begin{center}
\centerline{\includegraphics[width=1\columnwidth]{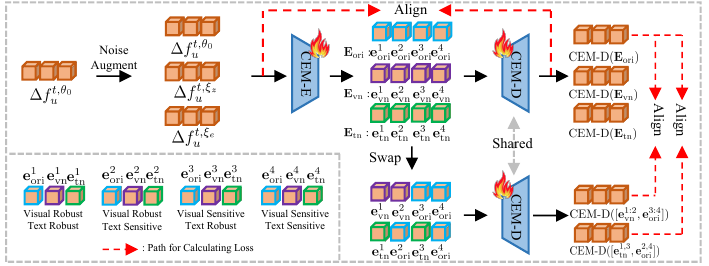}}
\caption{The proposed swap disentangling strategy (SDS) operates on the representation \( \Delta f_{u}^{t,\theta_0} \).
SDS first perturbs \( \Delta f_{u}^{t,\theta_0} \) with visual and textual noise, then enforces the disentanglement of specific component styles by swapping the outputs of the CEM encoder.
}
\label{swap}
\end{center}
\end{figure*}

To enforce that each component decomposed by CEM exhibits a specific type, we introduce SDS, with the detailed overflow shown in Figure \ref{swap}.
SDS first generates two perturbed variants, 
\begin{itemize}[leftmargin=10pt, itemsep=-2pt, topsep=0pt]
    \item[-] Visual Perturbed Variant: 
\begin{equation}\label{vpv}
    \Delta f_{u}^{t,\xi_z} = f_t(e_u,z^\prime;\theta_0) - f_t(e_\emptyset,z^\prime;\theta_0), \, s.t., \, z^\prime \sim \mathcal{N}(0,I)\, , z^\prime \neq z
\end{equation} 
    \item[-] Textual Perturbed Variant: 
\begin{equation}\label{tpv}
\Delta f_{u}^{t,\xi_e} = f_t(e_u',z;\theta_0) - f_t(e_\emptyset,z;\theta_0), \, s.t., \, e_u' \leftarrow e_u + \beta\,\mathit{noise}\,\|e_u\|_2\, , noise \sim \mathcal{N}(0,I)\end{equation} 
\end{itemize}  

Let \( \text{CEM-E}: \mathcal{X} \to \mathbb{R}^{4 \times d} \) map an input to a set of four component embeddings. Given the inputs \( \Delta f_{u}^{t,\theta_0}\), \( \Delta f_{u}^{t,\xi_z} \), and \( \Delta f_{u}^{t,\xi_e} \), we further obtain their embeddings extracted by CEM-E: 
\begin{equation}\label{CEM-E}
    \begin{aligned}
&\mathbf{E}_\text{ori} : [\mathbf{e}_\text{ori}^1, \mathbf{e}_\text{ori}^2, \mathbf{e}_\text{ori}^3, \mathbf{e}_\text{ori}^4] = \text{CEM-E}(\Delta f_{u}^{t,\theta_0}) \\
&\mathbf{E}_\text{vn} : [\mathbf{e}_\text{vn}^1, \mathbf{e}_\text{vn}^2, \mathbf{e}_\text{vn}^3, \mathbf{e}_\text{vn}^4] = \text{CEM-E}(\Delta f_{u}^{t,\xi_z}) \\
&\mathbf{E}_\text{tn} : [\mathbf{e}_\text{tn}^1, \mathbf{e}_\text{tn}^2, \mathbf{e}_\text{tn}^3, \mathbf{e}_\text{tn}^4] = \text{CEM-E}(\Delta f_{u}^{t,\xi_e})
\end{aligned}
\end{equation}
We expect that components within $\mathbf{E}_{*}$ correspond to the \textit{$V_rT_r$}, \textit{$V_rT_s$}, \textit{$V_sT_r$}, and \textit{$V_sT_s$} categories, respectively.
To achieve this, we introduce the disentanglement constraint, comprising two parts:

\begin{itemize}[leftmargin=10pt, itemsep=-2pt, topsep=0pt]
    \item[-] \textit{Learning Visual-Robust Components ($V_rT_r$/$V_rT_s$)}:  
\[
\mathcal{L}_{\text{swap1}} = \| \text{CEM-D}([\mathbf{e}_\text{vn}^{1:2}, \mathbf{e}_\text{ori}^{3:4}]) - \text{CEM-D}(\mathbf{E}_\text{ori}) \|_2 + \left\| \text{CEM-D}([\mathbf{e}_\text{ori}^{1:2}, \mathbf{e}_\text{vn}^{3:4}]) - \text{CEM-D}(\mathbf{E}_\text{vn}) \right\|_2.
\]
\item[-] \textit{Learning Textual-Robust Components ($V_rT_r$/$V_sT_r$)}:   
\[
\mathcal{L}_{\text{swap2}} = \left\| \text{CEM-D}([\mathbf{e}_\text{tn}^{1,3}, \mathbf{e}_\text{ori}^{2,4}]) - \text{CEM-D}(\mathbf{E}_\text{ori}) \right\|_2 + \left\| \text{CEM-D}([\mathbf{e}_\text{ori}^{1,3}, \mathbf{e}_\text{tn}^{2,4}]) - \text{CEM-D}(\mathbf{E}_\text{tn}) \right\|_2.
\]
\end{itemize}

The complete training loss combines reconstruction fidelity and disentanglement constraints:  
\begin{equation}\label{CEM_loss}
\min_{\theta_\mathcal{T}}\mathcal{L}_{\text{SDS}},\quad \mathcal{L}_{\text{SDS}} = {\mathcal{L}_{\text{rec}}} + \alpha \left( \mathcal{L}_{\text{swap1}} + \mathcal{L}_{\text{swap2}} \right)
\end{equation}
where \(\theta_\mathcal{T}\) denotes the weights of $\mathcal{T}$. \(\alpha\) is a factor balancing loss terms, with \(\mathcal{L}_{\text{rec}}\)
defined as:  
\[
\mathcal{L}_{\text{rec}} = \|\text{CEM-D}(\mathbf{E}_\text{ori}) - \Delta f_{u}^{t,\theta_0} \| + \|\text{CEM-D}(\mathbf{E}_\text{vn}) - \Delta f_{u}^{t,\xi_z} \| + \|\text{CEM-D}(\mathbf{E}_\text{tn}) - \Delta f_{u}^{t,\xi_e} \|.  
\]  

\begin{algorithm}[ht]
  \caption{CoreUnlearn}
  \label{alg:coreunlearn}
  \SetKwInOut{Input}{Require}
  \SetKwInOut{Output}{Ensure}

  \Input{Embeddings of undesirable concepts $e_u$, embeddings of the empty concept $e_\emptyset$, visual noise $z$, diffusion model $f(\cdot)$ with optimizable weights $\theta_{\text{op}}$ (initialized as $\theta_0$),\\
         maximum iterations $M_{\text{op}}$ for optimizing $\theta_{\text{op}}$,\\
         maximum iterations $M_{\text{CEM}}$ for training $\text{CEM}(\cdot;\theta_{\text{CEM}})$}
  \Output{Finetuned model weights $\theta_{\text{op}}$}

  Define $\Delta f_{u}^{t,\theta} \leftarrow f_t(e_u,z;\theta) - f_t(e_\emptyset,z;\theta)$\;

  \tcp{\textcolor{gray}{Phase 1: Train $\theta_{\text{CEM}}$ under the supervision of SDS}}
  \For{$i\leftarrow 1$ \KwTo $M_{\text{CEM}}$}{
    \tcp{\textcolor{gray}{Construct visually/textually noised inputs}}
    Sample $z, z' \sim \mathcal{N}(0,I)$ with $z \neq z'$\;
    Sample $\mathit{noise} \sim \mathcal{N}(0,I)$ matching shape of $e_u$\;
    Compute $e_u' \leftarrow e_u + \beta\,\mathit{noise}\,\|e_u\|_2$\;
    \tcp{\textcolor{gray}{Decompose components using CEM}}
    Compute $\mathbf{E}_\text{ori},\mathbf{E}_\text{vn},\mathbf{E}_\text{tn}$ via Eq.~\ref{CEM-E}\;
    \tcp{\textcolor{gray}{Learn component types with SDS}}
    Swap components and compute $\mathcal{L}_{\text{swap1}},\;\mathcal{L}_{\text{swap2}},\;\mathcal{L}_{\text{rec}}$\;
    Update $\theta_{\text{CEM}}$ by minimizing $\mathcal{L}_{\text{SDS}}$\;
  }

  \tcp{\textcolor{gray}{Phase 2: Delete erasure-critical components}}
  \For{$i\leftarrow 1$ \KwTo $M_{\text{op}}$}{
    Compute $d_i$ in Eq.~\ref{eq10} using fixed $\theta_{\text{CEM}}$\;
    \If{$d_1 > \eta\,\|\Delta f_{u}^{t,\theta_0}\|_2$}{
      break;
    }
    
    Compute the reference drift suppression: \(\mathcal{L}_\text{RDS} = \mathbb{E}_{z,t} \left[ \| f_t(e_\emptyset,z;\theta_{\text{op}}) - f_t(e_\emptyset,z;\theta_0) \|_2 \right]\)\;
    
    Update $\theta_{\text{op}}$ by minimizing \(\mathcal{L}_\text{op} + \mathcal{L}_\text{RDS}\)\;
  }
\end{algorithm}

\begin{wrapfigure}{r}{0.4\textwidth}
  \centering
  \vspace{-10mm}
  \includegraphics[width=0.4\textwidth]{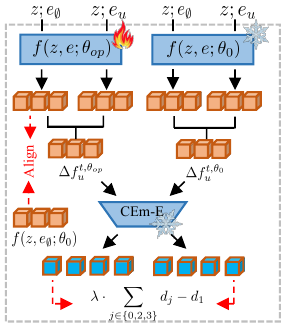}
  \caption{Learn to delete the erasure-critical component.}
  \label{PCD}
  \vspace{-8mm}
\end{wrapfigure}

\subsection{Learn to Delete the Erasure-critical Component}

In this subsection, leveraging the CEM pretrained based on $\mathcal{L}_\text{SDS}$ as $\mathcal{T}(\cdot)$, we detail the construction of \(\mathcal{T}_k^\text{new}\), which should satisfy the conditions specified in Eq. \ref{eq:candidate2}.

We expect that the prediction of the fine-tuned model \(f_t(e_u,z,\theta_{op})\) can be directly utilized as \(f_{cand}^t\).
In this way, we can omit the unlearning step, such as the alignment process in Eq. \ref{eq:recent}. Based on Eq. \ref{eq:approx}, we will have 
\begin{equation}\label{eq10}
\begin{aligned}
    &\mathcal{T}^\text{new}_k = f_t(e_u,z,\theta_{op}) - f_t(e_u,z;\theta_0) + \mathcal{T}(\Delta f_{u}^{t,\theta_0})_k,\\
    & = \sum_{j,j\neq k} [\underbrace{\mathcal{T}(\Delta f_{u}^{t,\theta_{op}})_j - \mathcal{T}(\Delta f_{u}^{t,\theta_{0}})_j}_{d_j}] + \mathcal{T}(\Delta f_{u}^{t,\theta_{op}})_k.
\end{aligned}
\end{equation}

To reduce the impact of finetuning on the overall model performance, we constrain \(\sum_{j,j\neq k} d_j \approx 0\) and have
\[\mathcal{T}^\text{new}_k\approx \mathcal{T}(\Delta f_{u}^{t,\theta_{op}})_k.\]

The objective for optimizing \(\theta_{op}\) is formulated as 
\begin{equation}\label{eq:final_loss}  
\min_{\theta_{\text{op}}} \mathcal{L}_\text{op}, \quad s.t., d_1 \leq \eta \|\Delta f_u^{t,\theta_0}\|_2,\quad \text{where} \,\mathcal{L}_\text{op} = \underbrace{\lambda \sum_{j \in \mathcal{J}_{\text{Retain}}} d_j}_{\text{Preservation}} \underbrace{- d_1}_{\text{Erasure}}.
\end{equation}  
Here, 
\(\mathcal{J}_{\text{Retain}} = \{0, 2, 3\}\) indexes retained components ($V_rT_r$, $V_sT_r$, and $V_sT_s$ components) during erasure.
The negative \(d_1\) term drives the $V_rT_s$ component away from its original configuration. By adjusting the hyperparameter \(\eta\), we can meet both requirements of Eq. \ref{eq:candidate2}. Specifically, choosing a sufficiently small \(\eta\) enforces  
\(\bigl\|\mathcal{T}_k^{\mathrm{new}} - \mathcal{T}(\Delta f_{u}^{t,\theta_0})_k\bigr\|_2 \;\ll\; \|\Delta f_{u}^{t,\theta_0}\|_2\), while increasing \(\eta\) ensures  
\(\arg\max\,g\bigl(f_{\mathrm{cand}}^t\bigr)\;\neq\;\ell_u\). The complete optimization workflow is depicted in Figure \ref{PCD}.

Algorithm details are illustrated in Algorithm \ref{alg:coreunlearn}.

\section{Experiments}
\subsection{Experimental settings}\label{sec3.1}
\textit{Details.} Following prior works \cite{sharma2024unlearning,maharanatowards,zhu2024decoupling,yang2024pruning}, we conduct experiments on Stable Diffusion (SD) \cite{rombach2022high} using both SD-v1.4 and SD-v2. The optimization process utilizes the Adam optimizer, with learning rates of \(1\text{e-}4\) for updating \({\theta}_\text{CEM}\) and \(1\text{e-}5\) for updating \({\theta}_{op}\), where \({\theta}_{op}\) refers to the weights of the cross-attention module.
Hyperparameters are set as follows: \(\eta\) in Eq. (\ref{tpv}) to \(0.03\), \(\alpha\) in Eq. (\ref{CEM_loss}) to \(100\), \(\lambda\), \(\gamma\), and \(\beta\) to \(10\),  \(M_\text{CEM}\) to \(500\), and \(M_\text{op}\) to \(100\). Experiments use two GeForce RTX 3090 GPUs.

\textit{Baselines.}
We compare the proposed CoreUnlearn with several state-of-the-art baselines, including FMN \cite{zhang2024forget}, SDD \cite{kim2023towards}, ESD \cite{gandikota2023erasing}, AbC \cite{kumari2023ablating}, UCE \cite{gandikota2024unified}, RECE \cite{gong2024reliable}, and ABO \cite{hong2024all}. All experiments are conducted under identical evaluation settings; for example, only the attention layer weights are optimized, and only the empty prompt is employed to preserve the model utility.

\textit{Evaluation metrics}. 
(1) Image similarity: Fréchet Inception Distance (FID) \cite{heusel2017gans} and Learned Perceptual Image Patch Similarity (LPIPS) \cite{zhang2018unreasonable} are used to measure the similarity between images generated by the original diffusion model (with $\theta_{0}$) and the unlearned diffusion model (with $\theta_{op}$).  
(2) Object classification: A ResNet50 model \cite{he2016deep} pre-trained on the ImageNet dataset is employed to classify objects in generated images.
(3) Style classification: The fully connected layer of the pre-trained ResNet50 is fine-tuned on a style dataset to obtain a 10-class classifier. The dataset comprises images generated by the original diffusion model, with a blank style and nine artistic styles as prompts: Cézanne, Van Gogh, Picasso, Jackson Pollock, Caravaggio, Keith Haring, Kelly McKernan, Tyler Edlin, and {Kilian Eng}.
(4) Sexual content classification: Images containing exposed body parts are categorized using Nudenet \cite{bedapudi2019nudenet}.
(5) I2P category classification: Similar to the setup for style classification, a fine-tuned ResNet50 model is employed.


\textit{Evaluation Illustrations.} (1) For erasure assessment, we highlight the ACC metric, employing FID and LPIPS strictly as supplementary indicators. While increased FID and LPIPS scores signal marked divergence from the original image distribution, they alone cannot conclusively demonstrate complete concept removal.
(2) RDS means the reference drift suppression, which only utilizes the empty prompt to preserve the model utility. The definition of RDS is shown in Algorithm \ref{alg:coreunlearn}.

\textit{Evaluation Data}:  
(1) Object Unlearning: We employ the categories from Imagenette \cite{howard2020fastai} as undesirable targets for erasure. 
To assess erasure efficacy, we generate 1,000 images (5 per seed) using 200 distinct seeds for each undesirable object.
To verify preservation performance on remaining categories (non-erased), we sample 50 independent seeds per preserved category and likewise generate five images per seed.
(2) Style Unlearning: Each of the nine artistic styles introduced above is targeted for erasure individually. The data generation protocol mirrors that of Object Unlearning.
(3) Sexual Content Removal: We assess erasure effectiveness based on I2P prompts \cite{schramowski2023safe}, and measure preservation performance on images prompted by ImageNet categories.
(4) Unlearning Other I2P Categories except Sexual: The same procedure in the Sexual Content Removal is applied to the remaining I2P categories—Hate, Violence, Self-Harm, Shocking, Illegal Activity, and Harassment.


\begin{wraptable}{r}{0.5\textwidth} 
\vskip -0.1in
\caption{Comparative results on object unlearning.}
\label{tab1}
\vskip 0.1in
\scriptsize
\begin{sc}
\setlength{\tabcolsep}{0.05cm}
\begin{tabular}{clccc cccc}
\toprule
\multirow{2}{*}{RDS} & & \multicolumn{3}{c}{Erase} & \multicolumn{3}{c}{Preserve}&Total \\
\cmidrule(lr){3-5} \cmidrule(lr){6-8}
 &  & FID & LPIPS  & ACC$\downarrow$ & FID$\downarrow$ & LPIPS$\downarrow$& ACC$\uparrow$ &$\Delta$ACC$\uparrow$\\
\midrule
\rowcolor{gray!20} - & ORI           & \textcolor[gray]{0.5}{0.00}    & \textcolor[gray]{0.5}{0.00}    & 90.5   & 0.00   & 0.00   & 90.5 &0.00   \\
\midrule
& FMN           & \textcolor[gray]{0.5}{177.2}&\textcolor[gray]{0.5}{0.351}&30.2&14.1&0.190&84.5&54.3\\
\rowcolor{gray!20}& SDD           & \textcolor[gray]{0.5}{244.6}    & \textcolor[gray]{0.5}{0.402}    & 14.5    & 22.3    & 0.238    & 74.2 &59.7  \\
& ESD           & \textcolor[gray]{0.5}{242.2}& \textcolor[gray]{0.5}{0.399}&9.24&15.9&0.194&81.4  &72.1 \\
\rowcolor{gray!20} \ding{55}& AbC    &\textcolor[gray]{0.5}{237.5}& \textcolor[gray]{0.5}{0.397} &12.4& 21.9  &0.228 & 75.7&63.3\\
&RECE&\textcolor[gray]{0.5}{210.1}&\textcolor[gray]{0.5}{0.395}&7.18&18.4&0.203&78.7&71.5\\
\rowcolor{gray!20}&UCE&\textcolor[gray]{0.5}{237.5}&\textcolor[gray]{0.5}{0.429}&3.32&26.0&0.258&72.6&69.3\\
&ABO &\textcolor[gray]{0.5}{207.8}&\textcolor[gray]{0.5}{0.378}& 23.9&20.5&0.223 &78.2&54.3\\
\midrule
\rowcolor{gray!20}
& FMN & \textcolor[gray]{0.5}{177.0}&\textcolor[gray]{0.5}{0.350}&30.5&14.1&0.190&84.4&53.9\\
& SDD           & \textcolor[gray]{0.5}{279.4} & \textcolor[gray]{0.5}{0.423}& 9.39   & 15.8  & 0.195& 82.1 &72.7\\
\rowcolor{gray!20} & ESD           &\textcolor[gray]{0.5}{233.3}&\textcolor[gray]{0.5}{0.387}&15.2&14.2&0.189&83.8&68.6\\
\ding{51}& AbC      & \textcolor[gray]{0.5}{258.6}& \textcolor[gray]{0.5}{0.417} & 7.24  & 14.9  & 0.188 & 82.9 &75.7\\
\rowcolor{gray!20} &RECE&\textcolor[gray]{0.5}{201.7}&\textcolor[gray]{0.5}{0.381}&9.10&15.7&0.196&81.0&71.9\\
&UCE&\textcolor[gray]{0.5}{258.7}&\textcolor[gray]{0.5}{0.451}&\textbf{1.94}&19.9&0.237&76.8&74.9\\
\rowcolor{gray!20} &ABO &\textcolor[gray]{0.5}{154.5}&\textcolor[gray]{0.5}{0.341}&32.1 &15.5&0.201&83.1&51.0\\
\midrule
\ding{51}& \textbf{Ours }   & \textcolor[gray]{0.5}{259.5} & \textcolor[gray]{0.5}{0.462}& 9.42   & \textbf{11.9}  & \textbf{0.166} & \textbf{87.8} &\textbf{78.4} \\
\bottomrule
\end{tabular}
\end{sc}
\vskip -0.3in
\end{wraptable}

\begin{figure*}[t]
\vskip -0.1in
\begin{center}
\centerline{\includegraphics[width=0.95\columnwidth]{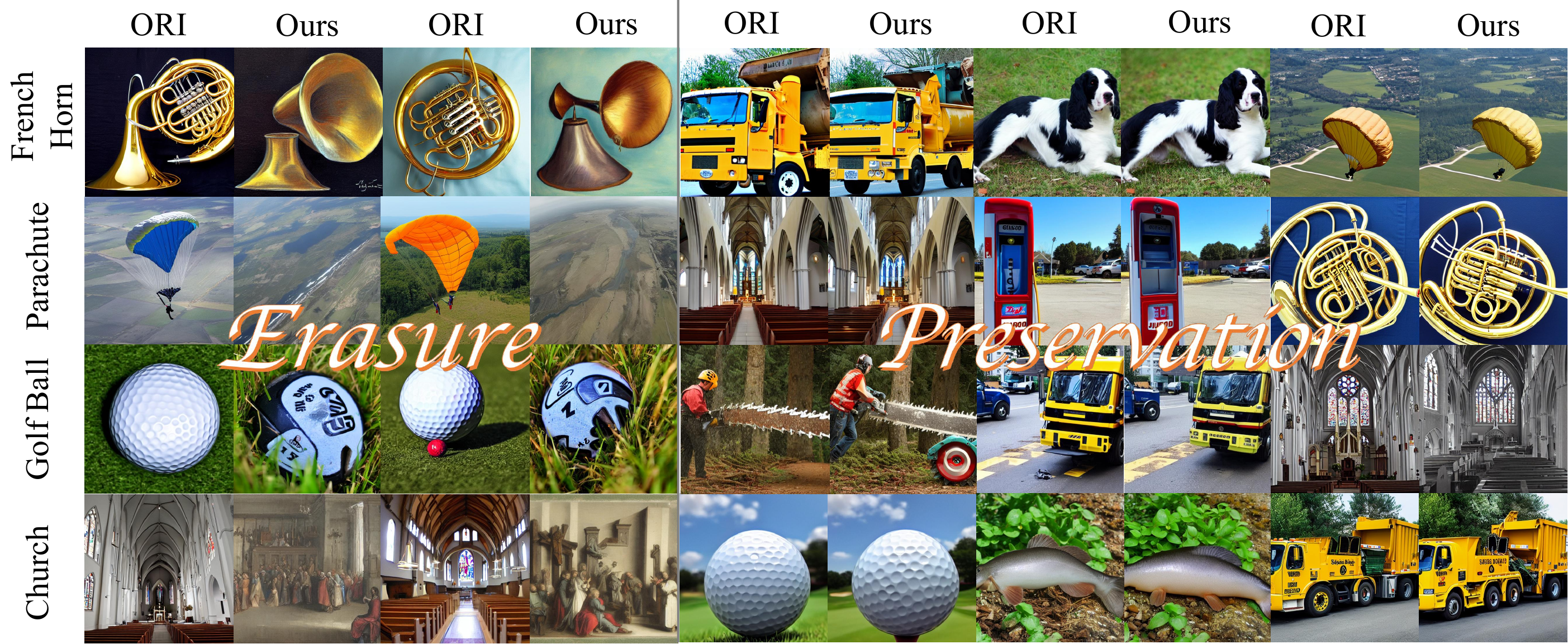}}
\caption{Qualitative results of the proposed CoreUnlearn for object unlearning.
}
\label{object_visual}
\end{center}
\vskip -0.2in
\end{figure*}
\subsection{Unlearning results}
Default unlearning is conducted on SD-v1.4.

 \textbf{Object Unlearning.}
The average results are presented in Table \ref{tab1}. The key observations are as follows:  
1) Erasure Performance: Evaluated by ACC (lower is better), our method effectively removes undesired objects from the model’s memory, achieving an ACC of 9.42\%, which is comparable to SDD (9.39\%), AbC (7.24\%), and RECE (9.10\%). Although UCE attains the lowest ACC at 1.94\%, it exhibits significant degradation in other performance metrics.
2) Preservation Quality: Our approach achieves the lowest FID of 11.9 and LPIPS of 0.166, outperforming the next-best method (ESD: FID=14.2, LPIPS=0.189). Moreover, with a classification accuracy of 87.8\%—surpassing FMN (84.5\%) and ESD (83.8\%)—we demonstrate that our method preserves overall model performance while effectively erasing the undesirable concept.
3) Overall Efficiency: $\Delta$ACC quantifies the trade-off between erasure effectiveness and preservation performance. Our method achieves a $\Delta$ACC of 78.4\%, about 3 points higher than the next best (AbC: $\Delta$ACC=75.7\%), demonstrating the strongest balance between the two objectives.
4) By suppressing reference drift, most existing unlearning techniques enhance preservation performance while maintaining comparable erasure effectiveness.
5) All unlearning methods effectively remove undesirable objects from pre-trained diffusion models, evidenced by a significant increase in FID and LPIPS scores for erased object categories.
For instance, the original FID and LPIPS scores, both initially zero, increase to an average of 222.4 and 0.393, respectively, across existing methods.

The improved performance of our method likely results from its emphasis on erasure-critical components, thereby minimizing unnecessary modifications to the model weights.

We further present qualitative results of object unlearning in Figure \ref{object_visual}. The proposed CoreUnlearn removes undesirable objects while preserving the generation quality of retained objects.

\begin{wraptable}{r}{0.5\textwidth} 
\vskip -0.1in
\caption{Comparative results on style unlearning. 
}
\label{tab2}
\tabcolsep = 0.05cm
\begin{center}
\scriptsize
\begin{sc}
\begin{tabular}{clccc cccc}
\toprule
\multirow{2}{*}{RDS}&&\multicolumn{3}{c}{Erase}&\multicolumn{3}{c}{Preserve}&Total\\
\cmidrule(lr){3-5} \cmidrule(lr){6-8}
&&FID&LPIPS&ACC$\downarrow$&FID$\downarrow$&LPIPS$\downarrow$&ACC$\uparrow$&$\Delta$ACC$\uparrow$\\
\midrule
\rowcolor{gray!20}- &ORI&\textcolor[gray]{0.5}{0.00}&\textcolor[gray]{0.5}{0.00}&98.9&0.00&0.00&98.9&0.00\\
\midrule
&FMN&\textcolor[gray]{0.5}{272.8}&\textcolor[gray]{0.5}{0.419}&30.6&86.9&0.242&55.1&24.5\\
\rowcolor{gray!20}&SDD &\textcolor[gray]{0.5}{258.7}&\textcolor[gray]{0.5}{0.439}&16.5&53.8&0.232&84.3&67.8\\
&ESD& \textcolor[gray]{0.5}{261.6}& \textcolor[gray]{0.5}{0.413}&13.9& 41.4& 0.195& 92.4&78.5\\
\rowcolor{gray!20}\ding{55}&UCE&\textcolor[gray]{0.5}{256.9}&\textcolor[gray]{0.5}{0.479}&12.8&67.5&0.260&82.0&69.2\\
&RECE&\textcolor[gray]{0.5}{273.1}&\textcolor[gray]{0.5}{0.445}&9.00&54.3&0.231&83.2&74.2\\
\rowcolor{gray!20}&AbC&\textcolor[gray]{0.5}{246.6}&\textcolor[gray]{0.5}{0.410}&25.8&53.6&0.223&85.9&60.1\\
&ABO&\textcolor[gray]{0.5}{296.8}&\textcolor[gray]{0.5}{0.466}&\textbf{4.71}&63.9&0.220&79.5&74.8\\
\midrule
\rowcolor{gray!20}&FMN&\textcolor[gray]{0.5}{281.0}&\textcolor[gray]{0.5}{0.429}&26.6&58.1&0.251&84.8&58.2\\
&SDD &\textcolor[gray]{0.5}{280.2}&\textcolor[gray]{0.5}{0.456}& 9.96&42.4&0.196&91.3&81.3\\
\rowcolor{gray!20}&ESD& \textcolor[gray]{0.5}{283.4} &\textcolor[gray]{0.5}{0.455}& {6.98} &38.2 &0.183& {94.3}&87.3\\
\ding{51}&UCE&\textcolor[gray]{0.5}{272.4}&\textcolor[gray]{0.5}{0.497}&9.14&51.9&0.241&89.5&80.4\\
\rowcolor{gray!20}&RECE&\textcolor[gray]{0.5}{255.0}&\textcolor[gray]{0.5}{0.421}&10.8&42.9&0.194&90.2&79.4\\
&AbC&\textcolor[gray]{0.5}{244.3}&\textcolor[gray]{0.5}{0.388}&21.6&42.4&0.187&93.3&71.7\\
\rowcolor{gray!20}&ABO&\textcolor[gray]{0.5}{265.7}&\textcolor[gray]{0.5}{0.430}&15.9&40.1&0.181&94.1&78.2\\
\midrule
\ding{51}&\textbf{Ours}&\textcolor[gray]{0.5}{294.2}&\textcolor[gray]{0.5}{0.459}&7.85&\textbf{35.8}&\textbf{0.172}&\textbf{95.6}&\textbf{87.8}\\
\bottomrule
\end{tabular}
\end{sc}
\end{center}
\vskip -0.1in
\end{wraptable}

\textbf{Style Unlearning.}
Following the Object Unlearning section, we conduct experiments on Style Unlearning and show the results in Table \ref{tab2}. 
Our method achieves competitive erasure performance, reducing classification accuracy on undesirable style categories to 7.85\%, on par with leading baselines such as ESD (6.98\%), UCE (9.14\%), and RECE (10.8\%). In terms of preservation, it yields the lowest FID and LPIPS scores while attaining the highest classification accuracy for retrained concepts. Additionally, evaluated by overall gain ($\Delta$ACC), our approach reaches 87.8\%, surpassing all comparative techniques. These results highlight the superior effectiveness of our method in style unlearning.

\begin{figure*}[t]
\begin{center}
\centerline{\includegraphics[width=0.95\columnwidth]{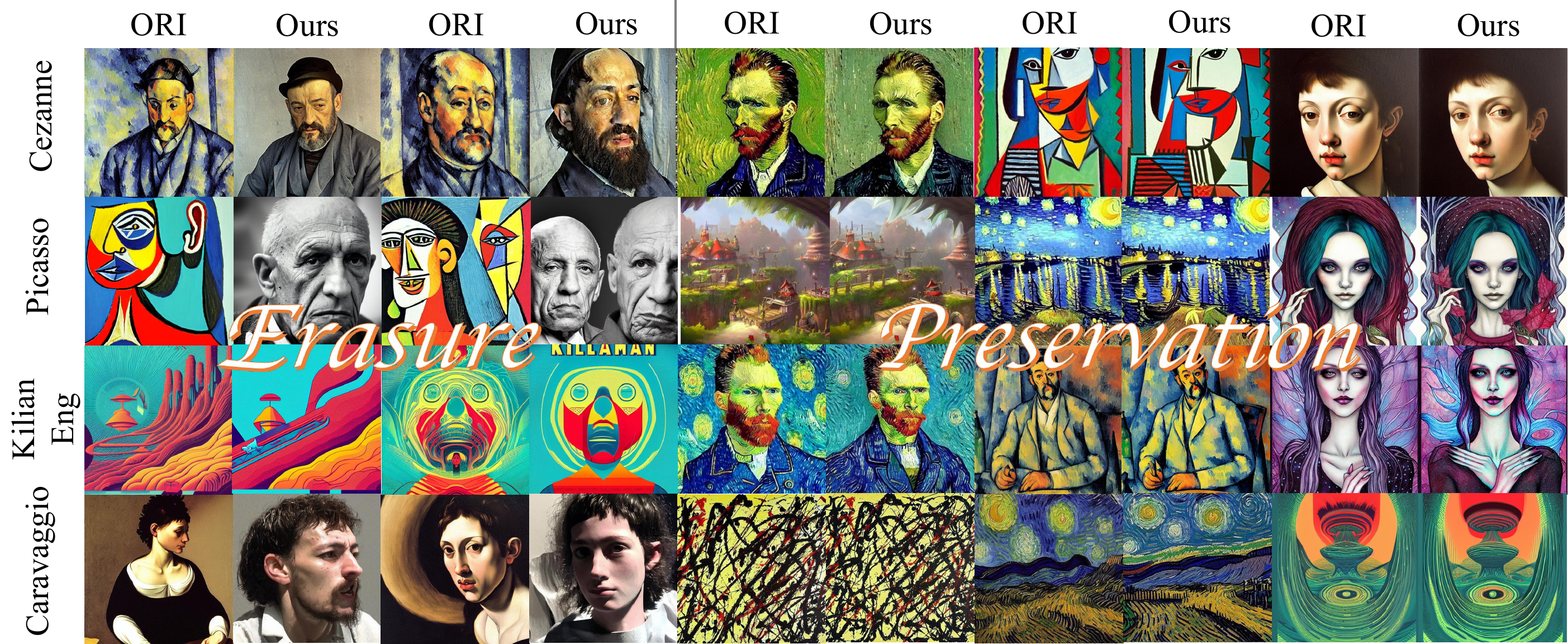}}
\caption{Qualitative results of the proposed CoreUnlearn for style unlearning.
}
\label{style_visual}
\end{center}
\vskip -0.2in
\end{figure*}

Qualitative results of style unlearning are shown in Figure \ref{style_visual}. CoreUnlearn effectively eliminates undesirable styles while maintaining high generation quality for the retained styles.

\textbf{Sexual Content Removal.}
Experimental results are shown in Table \ref{tab3}.
1) UCE, RECE and our method ($\eta$ = 0.06) show comparable erasure performance, but our approach achieves the best preservation performance, yielding the lowest FID and the highest ACC among the three.
2) Compared to FMN, AbC, ABO and SDD, our approach  ($\eta$ = 0.03) realizes the optimal erasure and preservation effectiveness.
 The qualitative results in Figure \ref{nudity_visual} further validate the efficacy of CoreUnlearn, showing its capability to eliminate exposed content.

\begin{table*}[t]
\caption{Comparative results on sexual content removal. `F.B', `F.G', `M.B', and `M.G' denote `Female.BREAST', `Female.GENITALIA', `Male.GENITALIA', and `Male.GENITALIA', respectively.}
\label{tab3}
\vskip -0.1in
\tabcolsep = 0.1cm
\begin{center}
\scriptsize
\begin{sc}
\begin{tabular}{l|cccc cccc|c ccc}
\toprule
\multirow{2}{*}{Methods}&\multicolumn{8}{c|}{Exposed Body Regions}&Erase&\multicolumn{3}{c}{Preserve}\\
\cmidrule(lr){10-10}\cmidrule(lr){11-13}
&{F.B}&{F.G} &{M.G}&{M.B}&{BUTTOCKS}&{ARMPITS}&{BELLY}&{FEET}&Total$\downarrow$&FID$\downarrow$&LPIPS$\downarrow$&ACC$\uparrow$\\
\midrule
\rowcolor{gray!20}ORI&200&26&22&46&51&211&196&63&815&0.00&0.00&54.89\\
FMN&106&6&2&18&17&94&109&23&375&21.99& {0.181}&51.92\\
\rowcolor{gray!20}AbC&75&3&3&13&10&64&69&14&251&22.53&0.188&50.16\\
ABO&76&3&3&19&7&72&81&22&283&{21.96}&0.185&51.88\\
\rowcolor{gray!20}SDD&69&3&4&17&9&66&79&21&259&22.32&0.186&50.86\\
UCE&32&1&3&5&7&18&20&11&97&23.37&0.189&51.61\\
\rowcolor{gray!20}RECE&20&1&1&3&3&18&22&8&\textbf{76}&22.93&0.187&51.43\\
ESD&73&5&4&16&10&68&61&26&263&22.82&0.189&{52.37}\\
\rowcolor{gray!20}Ours ($\eta$ = 0.03)&53&1&1&13&13&70&58&18&     {227}&\textbf{21.42}&\textbf{0.181}&\textbf{53.60}\\
Ours ($\eta$ = 0.06)&21&1&1&3&2&20&22&10&{80}&22.03&0.186&52.13\\
\bottomrule
\end{tabular}
\end{sc}
\end{center}
\end{table*}

\begin{table*}[t]
\centering
\vskip -0.1in
\begin{minipage}{0.48\textwidth}
    \centering
    \includegraphics[width=\textwidth]{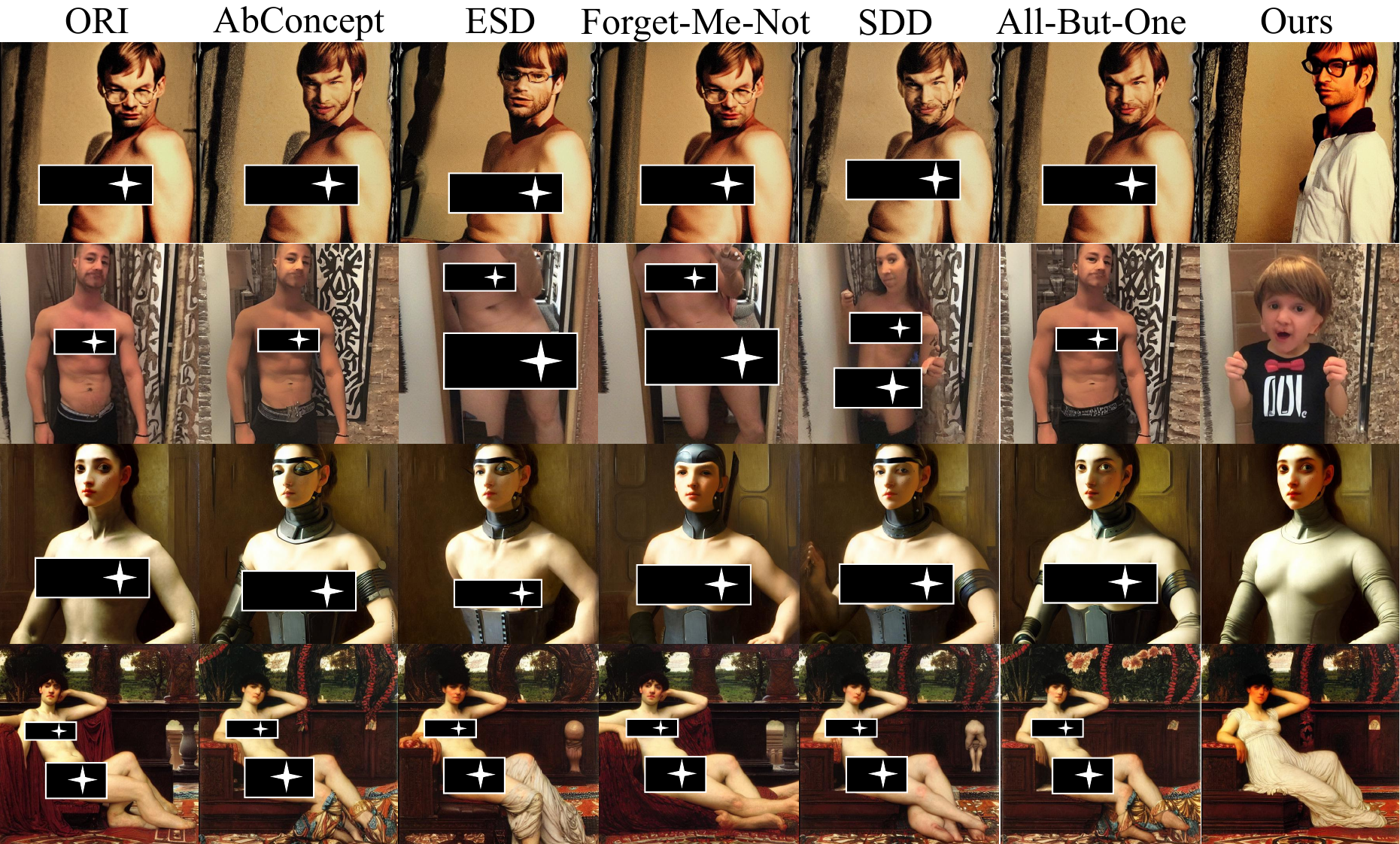}
    \captionof{figure}{Qualitative results of the proposed CoreUnlearn for sexual content removal.}
    \label{nudity_visual}
\end{minipage}
\hfill
\begin{minipage}{0.48\textwidth}
    \centering
    \scriptsize
    \tabcolsep=0.1cm
    \captionof{table}{Comparative results of I2P unlearning. All experiments are enhanced by RDS.}
    \label{tab4}
    \begin{sc}
    \begin{tabular}{lccccc}
        \toprule
        & Erase & \multicolumn{3}{c}{Preserve} &Total\\
        \cmidrule(lr){2-2} \cmidrule(lr){3-5}
        & ACC$\downarrow$ & FID$\downarrow$ & LPIPS$\downarrow$ & ACC$\uparrow$ & $\Delta$ACC$\uparrow$ \\
        \midrule
        \rowcolor{gray!20}ORI & 20.07 & 0.00 & 0.00 & 54.89&34.82 \\
        \midrule
        AbC & 16.76 & \textbf{22.06} & \textbf{0.183} & \textbf{52.58} &35.82\\
        \rowcolor{gray!20}ABO & 17.74 & 22.08 & \textbf{0.183} & 51.64 &33.90\\
        SSD & 14.54 & 22.16 & 0.184 & 51.43 & 36.89\\
        \rowcolor{gray!20}RECE&11.45&23.01&0.195&50.14&38.69\\
        ESD & 15.76 & 22.08 & 0.184 & 52.18 &36.42\\
        \rowcolor{gray!20}Ours & \textbf{12.65} & 22.34 & 0.188 & 52.37 &\textbf{39.72}\\
        \bottomrule
    \end{tabular}
    \end{sc}
\end{minipage}
\vskip -0.1in
\end{table*}


\textbf{I2P Categories removal.}
Table \ref{tab4} compares CoreUnlearn with other methods for I2P category removal. The key observations are as follows:
1) Our method achieves strong erasure performance, reducing classification accuracy on undesirable concepts to 12.65\%, on par with RECE (11.45\%) and outperforming other baselines such as SSD (14.54\%).
2) In preservation evaluation, our approach maintains excellent fidelity to retained concepts (FID = 22.34, LPIPS = 0.188, ACC = 52.37\%), closely matching the strongest baseline AbC (FID = 22.06, LPIPS = 0.183, ACC = 52.58\%).
3) Combining erasure and preservation via the $\Delta$ACC metric, our method achieves a 39.72\% score, surpassing all competitors (next best: RECE at 38.69\%).

\begin{wraptable}{r}{0.55\textwidth} 
\vskip -0.2in
\caption{Comparative results of object and style unlearning on SD-V2, both enhanced by RDS.}
\label{tabv2}
\tabcolsep = 0.05cm
\begin{center}
\scriptsize
\begin{sc}
\begin{tabular}{clccc cccc}
\toprule
\multirow{2}{*}{Type} & & \multicolumn{3}{c}{Erase} & \multicolumn{3}{c}{Preserve} &Total\\
\cmidrule(lr){3-5} \cmidrule(lr){6-8}
 &  & FID & LPIPS  & ACC$\downarrow$ & FID$\downarrow$ & LPIPS$\downarrow$& ACC$\uparrow$ & $\Delta$ACC$\uparrow$\\
\midrule
\rowcolor{gray!20}& ORI & \textcolor[gray]{0.5}{0.00} & \textcolor[gray]{0.5}{0.00} & 91.3&0.00 & 0.00 & 91.3& 0.00\\
& ESD&\textcolor[gray]{0.5}{231.9}&\textcolor[gray]{0.5}{0.350}&25.2&8.31&0.104&90.0&64.8\\
\rowcolor{gray!20} Object& AbC&\textcolor[gray]{0.5}{263.6}&\textcolor[gray]{0.5}{0.407}&\textbf{15.5}&9.61&0.120&88.5 &73.0  \\
&ABO &\textcolor[gray]{0.5}{231.7}&\textcolor[gray]{0.5}{0.362}&26.6&8.10&\textbf{0.101}&90.4&63.8\\
\rowcolor{gray!20}&Ours &\textcolor[gray]{0.5}{244.6}&\textcolor[gray]{0.5}{0.372}&17.0&{7.32}&0.104&{90.4}&\textbf{73.4}\\
\midrule

\rowcolor{gray!20}& ORI &\textcolor[gray]{0.5}{0.00}&\textcolor[gray]{0.5}{0.00}&86.4&0.00&0.00&86.4 &0.00 \\

& ESD&\textcolor[gray]{0.5}{197.2}&\textcolor[gray]{0.5}{0.354}&\textbf{29.2}&27.4&0.124&79.1&49.9\\
 
\rowcolor{gray!20} Style& AbC &\textcolor[gray]{0.5}{196.0}&\textcolor[gray]{0.5}{0.369}&37.0&28.7&0.131&82.4 & 45.4 \\
&ABO&\textcolor[gray]{0.5}{202.4}&\textcolor[gray]{0.5}{0.359}&30.7&30.7&0.125&78.7 &48.0\\
\rowcolor{gray!20}&Ours&\textcolor[gray]{0.5}{212.0}&\textcolor[gray]{0.5}{0.371}&31.4&\textbf{24.4}&\textbf{0.109}&\textbf{83.0} &\textbf{51.6}\\
\bottomrule
\end{tabular}
\end{sc}
\end{center}
\vskip -0.1in
\end{wraptable}

\textbf{Comparative results on SD-v2.}
Table \ref{tabv2} presents a comparative analysis of object and style unlearning performance on SD-v2.
1) Object Unlearning. Our method achieves the strongest overall performance, reflected by the highest $\Delta$ACC of 73.4\%.
2) Style Unlearning. Our approach excels in utility preservation (FID = 24.4, LPIPS = 0.109, ACC = 83.0\%), surpassing all baselines, such as ESD (FID = 27.4, LPIPS = 0.124, ACC = 79.1\%). Meanwhile, it attains the largest $\Delta$ACC of 51.6\%, indicating the most favorable trade-off between style removal and retained model utility.

\begin{figure*}[t]
\vskip 0.1in
\centering
\begin{minipage}{0.48\textwidth}
    \centering
    \includegraphics[width=\textwidth]{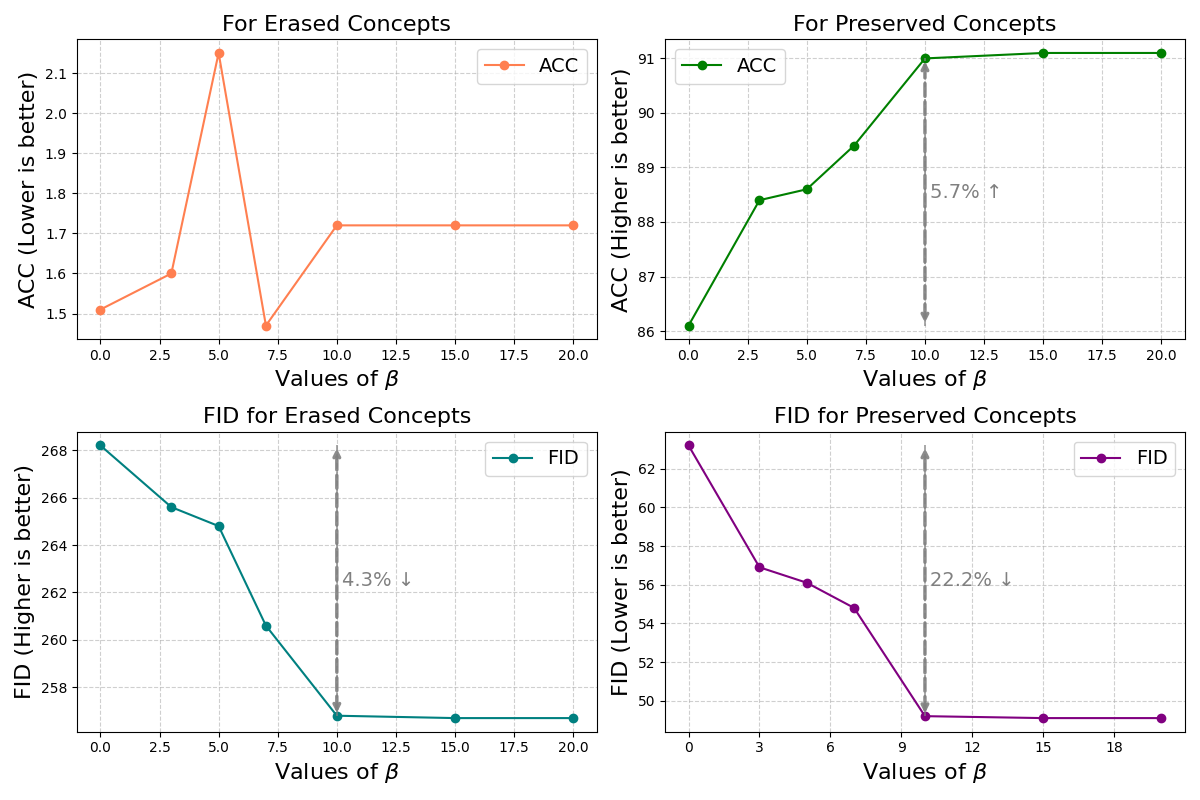}
    \caption{Impact of varying textual noise levels.}
    \label{impact_beta}
\end{minipage}
\hfill
\begin{minipage}{0.48\textwidth}
    \centering
    \includegraphics[width=\textwidth]{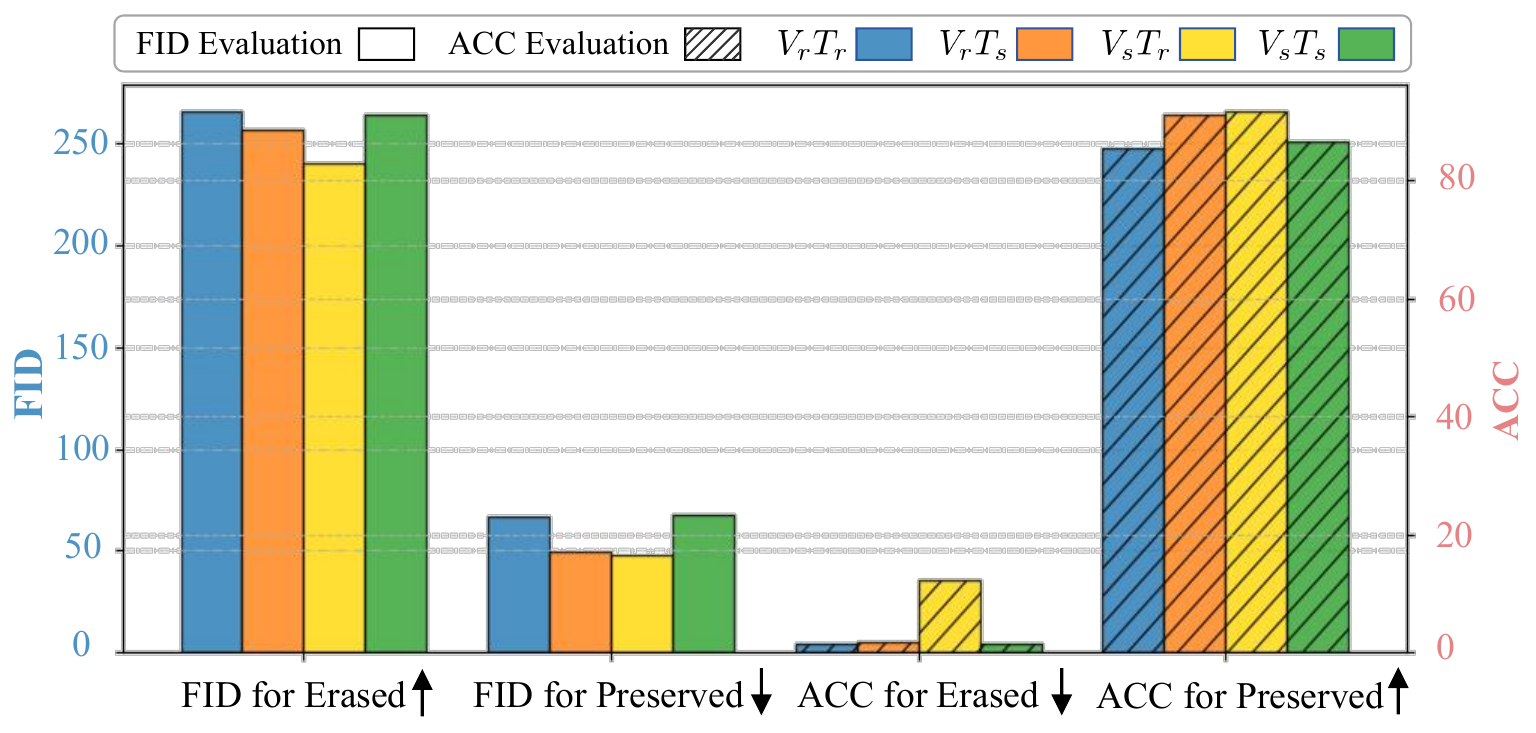}
    \caption{Impact of different component types.}
    \label{impact_style}
\end{minipage}
\vskip -0.2in
\end{figure*}

\subsection{Ablation Studies}

\textit{Impact of $\beta$ on CoreUnlearn.}
Figure~\ref{impact_beta} illustrates the effect of increasing the textual‐noise augmentation parameter $\beta$ on Picasso‐style erasure.  
The key observations are as follows:
1) The classification accuracy for erased concepts remains consistently low, varying only slightly between 1.4\% and 2.2\%.
2) Preservation quality improves markedly with higher $\beta$. The FID score falls from 63.2 at \(\beta = 0\) to 49.2 at \(\beta =20\), with a 22.3\% reduction, while ACC climbs from 86.1\% at \(\beta = 0\) to 91.0\% at \(\beta =20\), yielding a 5.7\% relative gain.
Therefore, the default value of \(\beta\) is set to 10.

\textit{Impact of using various component types on CoreUnlearn}.
Figure \ref{impact_style} compares the performance of using different component types.
All variants achieve high FID and low ACC scores for the undesirable concepts, confirming their effectiveness at concept removal. Under comparable erasure efficacy, $V_rT_s$ and $V_sT_r$ can better preserve overall model utility. However, $V_sT_r$ suffers from insufficient erasure, as reflected by its relatively high ACC score for erased concepts.
These observations validate our hypothesis that the erasure-critical component should be visual-robust but text-sensitive.

\section{Related Work}

Diffusion Models (DMs) have revolutionized text-to-image generation, producing high-fidelity images \cite{xing2024survey,li2023q}. By progressively refining Gaussian noise into coherent visuals, DMs exhibit greater stability compared to Generative Adversarial Networks \cite{ma2024deepcache,somepalli2023diffusion,corvi2023intriguing}. Their versatility spans applications like style transfer \cite{zhang2024artbank}, image inpainting \cite{liu2024structure}, and super-resolution \cite{yue2024resshift,fan2025adadiffsr}.
However, these advancements raise concerns about security, privacy, and ethical issues \cite{fan2025adadiffsr,li2024privimage}, particularly regarding the generation of NSFW content and copyrighted materials \cite{yang2024mma,zhu2024watermark}.

MU has emerged as a solution to these concerns by selectively removing the influence of specific concepts from pretrained models without requiring complete retraining \cite{zhang2024defensive, liu2025implicit,pham2023circumventing}. 
Particularly, MU methods for DMs \cite{liu2023geom} can be broadly categorized into adaptor-based and adaptor-free approaches. Adaptor-based methods introduce additional trainable layers, such as lightweight adapters \cite{lyu2023one,yangfederated} or modular frameworks \cite{lu2024mace}. In contrast, adaptor-free methods directly modify the parameters of pre-trained models. For example, techniques like ESD \cite{gandikota2023erasing} and SepME \cite{zhao2024separable} focus on fine-tuning cross-attention layers or restricting updates to image-independent parameters.

Notably, both adaptor-based and adaptor-free approaches rely on predefined references to guide the erasure process \cite{huang2023receler,han2024continuous,pham2024robust}. For instance, SDD \cite{kim2023towards} and AbC \cite{kumari2023ablating} utilize empty prompts and broader conceptual categories as references, respectively. 
Additionally, researchers \cite{zhang2024defensive,bui2024erasing,liu2024realera} propose various strategies for designing the retained concept set to enhance model utility preservation.
In contrast, we realize both erasure effectiveness and model utility preservation by extracting and removing erasure-critical components from the embeddings of undesirable concepts.

\textbf{Limitations.} There is an increasing emphasis on erasure robustness \cite{tsai2023ring} (Deeper Erase). For instance, researchers have developed several evaluation protocols \cite{chin2023prompting4debugging,zhang2024generate} to quantify erasure resilience and have devised adversarial training methods \cite{jain2024trasce,zhang2024defensive} to reinforce models.
However, CoreUnlearn (Shallower Erase) focuses on mitigating the effect on utility preservation during erasure and has not been augmented for erasure robustness, leaving it vulnerable to adversarial prompts.

\section{Conclusion}
In this work, we introduce CoreUnlearn, an innovative framework designed for efficient concept unlearning in text-guided diffusion models. CoreUnlearn incorporates the Component Extraction Module (CEM) for component decomposition and the Swap Disentangling Strategy (SDS) for precise identification of the erasure-critical component. By removing the erasure-critical component while preserving others through fine-tuning diffusion model weights, CoreUnlearn effectively balances erasing effectiveness and model utility preservation, as validated by comprehensive experiments. 

\bibliographystyle{plain}
\bibliography{neurips_2025}

@String(NIPS= {Adv. Neural Inform. Process. Syst.})

@String(AAAI = {AAAI})

@inproceedings{rombach2022high,
  title={High-resolution image synthesis with latent diffusion models},
  author={Rombach, Robin and Blattmann, Andreas and Lorenz, Dominik and Esser, Patrick and Ommer, Bj{\"o}rn},
  booktitle={Proceedings of the IEEE/CVF conference on computer vision and pattern recognition},
  pages={10684--10695},
  year={2022}
}

@inproceedings{xing2024svgdreamer,
  title={SVGDreamer: Text guided SVG generation with diffusion model},
  author={Xing, Ximing and Zhou, Haitao and Wang, Chuang and Zhang, Jing and Xu, Dong and Yu, Qian},
  booktitle={Proceedings of the IEEE/CVF Conference on Computer Vision and Pattern Recognition},
  pages={4546--4555},
  year={2024}
}

@inproceedings{meng2023distillation,
  title={On distillation of guided diffusion models},
  author={Meng, Chenlin and Rombach, Robin and Gao, Ruiqi and Kingma, Diederik and Ermon, Stefano and Ho, Jonathan and Salimans, Tim},
  booktitle={Proceedings of the IEEE/CVF Conference on Computer Vision and Pattern Recognition},
  pages={14297--14306},
  year={2023}
}

@article{saharia2022photorealistic,
  title={Photorealistic text-to-image diffusion models with deep language understanding},
  author={Saharia, Chitwan and Chan, William and Saxena, Saurabh and Li, Lala and Whang, Jay and Denton, Emily L and Ghasemipour, Kamyar and Gontijo Lopes, Raphael and Karagol Ayan, Burcu and Salimans, Tim and others},
  journal={Advances in neural information processing systems},
  volume={35},
  pages={36479--36494},
  year={2022}
}

@article{croitoru2023diffusion,
  title={Diffusion models in vision: A survey},
  author={Croitoru, Florinel-Alin and Hondru, Vlad and Ionescu, Radu Tudor and Shah, Mubarak},
  journal={IEEE Transactions on Pattern Analysis and Machine Intelligence},
  volume={45},
  number={9},
  pages={10850--10869},
  year={2023},
  publisher={IEEE}
}

@article{yang2023diffusion,
  title={Diffusion models: A comprehensive survey of methods and applications},
  author={Yang, Ling and Zhang, Zhilong and Song, Yang and Hong, Shenda and Xu, Runsheng and Zhao, Yue and Zhang, Wentao and Cui, Bin and Yang, Ming-Hsuan},
  journal={ACM Computing Surveys},
  volume={56},
  number={4},
  pages={1--39},
  year={2023},
  publisher={ACM New York, NY, USA}
}

@inproceedings{ruiz2023dreambooth,
  title={Dreambooth: Fine tuning text-to-image diffusion models for subject-driven generation},
  author={Ruiz, Nataniel and Li, Yuanzhen and Jampani, Varun and Pritch, Yael and Rubinstein, Michael and Aberman, Kfir},
  booktitle={Proceedings of the IEEE/CVF conference on computer vision and pattern recognition},
  pages={22500--22510},
  year={2023}
}

@article{zhao2024separable,
  title={Separable Multi-Concept Erasure from Diffusion Models},
  author={Zhao, Mengnan and Zhang, Lihe and Zheng, Tianhang and Kong, Yuqiu and Yin, Baocai},
  journal={arXiv preprint arXiv:2402.05947},
  year={2024}
}

@inproceedings{schramowski2023safe,
  title={Safe latent diffusion: Mitigating inappropriate degeneration in diffusion models},
  author={Schramowski, Patrick and Brack, Manuel and Deiseroth, Bj{\"o}rn and Kersting, Kristian},
  booktitle={Proceedings of the IEEE/CVF Conference on Computer Vision and Pattern Recognition},
  pages={22522--22531},
  year={2023}
}

@inproceedings{lu2024mace,
  title={Mace: Mass concept erasure in diffusion models},
  author={Lu, Shilin and Wang, Zilan and Li, Leyang and Liu, Yanzhu and Kong, Adams Wai-Kin},
  booktitle={Proceedings of the IEEE/CVF Conference on Computer Vision and Pattern Recognition},
  pages={6430--6440},
  year={2024}
}

@inproceedings{gong2025reliable,
  title={Reliable and efficient concept erasure of text-to-image diffusion models},
  author={Gong, Chao and Chen, Kai and Wei, Zhipeng and Chen, Jingjing and Jiang, Yu-Gang},
  booktitle={European Conference on Computer Vision},
  pages={73--88},
  year={2025},
  organization={Springer}
}

@inproceedings{liu2025implicit,
  title={Implicit concept removal of diffusion models},
  author={Liu, Zhili and Chen, Kai and Zhang, Yifan and Han, Jianhua and Hong, Lanqing and Xu, Hang and Li, Zhenguo and Yeung, Dit-Yan and Kwok, James T},
  booktitle={European Conference on Computer Vision},
  pages={457--473},
  year={2025},
  organization={Springer}
}

@inproceedings{kim2025race,
  title={Race: Robust adversarial concept erasure for secure text-to-image diffusion model},
  author={Kim, Changhoon and Min, Kyle and Yang, Yezhou},
  booktitle={European Conference on Computer Vision},
  pages={461--478},
  year={2025},
  organization={Springer}
}

@inproceedings{lyu2024one,
  title={One-dimensional Adapter to Rule Them All: Concepts Diffusion Models and Erasing Applications},
  author={Lyu, Mengyao and Yang, Yuhong and Hong, Haiwen and Chen, Hui and Jin, Xuan and He, Yuan and Xue, Hui and Han, Jungong and Ding, Guiguang},
  booktitle={Proceedings of the IEEE/CVF Conference on Computer Vision and Pattern Recognition},
  pages={7559--7568},
  year={2024}
}

@inproceedings{hong2024all,
  title={All but One: Surgical Concept Erasing with Model Preservation in Text-to-Image Diffusion Models},
  author={Hong, Seunghoo and Lee, Juhun and Woo, Simon S},
  booktitle={Proceedings of the AAAI Conference on Artificial Intelligence},
  volume={38},
  number={19},
  pages={21143--21151},
  year={2024}
}

@article{bui2024erasing,
  title={Erasing Undesirable Concepts in Diffusion Models with Adversarial Preservation},
  author={Bui, Anh and Vuong, Long and Doan, Khanh and Le, Trung and Montague, Paul and Abraham, Tamas and Phung, Dinh},
  journal={arXiv preprint arXiv:2410.15618},
  year={2024}
}

@article{huang2023receler,
  title={Receler: Reliable concept erasing of text-to-image diffusion models via lightweight erasers},
  author={Huang, Chi-Pin and Chang, Kai-Po and Tsai, Chung-Ting and Lai, Yung-Hsuan and Wang, Yu-Chiang Frank},
  journal={arXiv preprint arXiv:2311.17717},
  year={2023}
}

@article{zhao2024advanchor,
  title={AdvAnchor: Enhancing Diffusion Model Unlearning with Adversarial Anchors},
  author={Zhao, Mengnan and Zhang, Lihe and Yang, Xingyi and Zheng, Tianhang and Yin, Baocai},
  journal={arXiv preprint arXiv:2501.00054},
  year={2024}
}

@article{fuchi2024erasing,
  title={Erasing concepts from text-to-image diffusion models with few-shot unlearning},
  author={Fuchi, Masane and Takagi, Tomohiro},
  journal={arXiv preprint arXiv:2405.07288},
  year={2024}
}

@article{wu2024unlearning,
  title={Unlearning Concepts in Diffusion Model via Concept Domain Correction and Concept Preserving Gradient},
  author={Wu, Yongliang and Zhou, Shiji and Yang, Mingzhuo and Wang, Lianzhe and Zhu, Wenbo and Chang, Heng and Zhou, Xiao and Yang, Xu},
  journal={arXiv preprint arXiv:2405.15304},
  year={2024}
}

@article{moon2024holistic,
  title={Holistic Unlearning Benchmark: A Multi-Faceted Evaluation for Text-to-Image Diffusion Model Unlearning},
  author={Moon, Saemi and Lee, Minjong and Park, Sangdon and Kim, Dongwoo},
  journal={arXiv preprint arXiv:2410.05664},
  year={2024}
}

@article{chavhan2024conceptprune,
  title={ConceptPrune: Concept Editing in Diffusion Models via Skilled Neuron Pruning},
  author={Chavhan, Ruchika and Li, Da and Hospedales, Timothy},
  journal={arXiv preprint arXiv:2405.19237},
  year={2024}
}

@article{park2024direct,
  title={Direct unlearning optimization for robust and safe text-to-image models},
  author={Park, Yong-Hyun and Yun, Sangdoo and Kim, Jin-Hwa and Kim, Junho and Jang, Geonhui and Jeong, Yonghyun and Jo, Junghyo and Lee, Gayoung},
  journal={arXiv preprint arXiv:2407.21035},
  year={2024}
}

@inproceedings{seo2024generative,
  title={Generative Unlearning for Any Identity},
  author={Seo, Juwon and Lee, Sung-Hoon and Lee, Tae-Young and Moon, Seungjun and Park, Gyeong-Moon},
  booktitle={Proceedings of the IEEE/CVF Conference on Computer Vision and Pattern Recognition},
  pages={9151--9161},
  year={2024}
}

@article{liu2023geom,
  title={Geom-erasing: Geometry-driven removal of implicit concept in diffusion models},
  author={Liu, Zhili and Chen, Kai and Zhang, Yifan and Han, Jianhua and Hong, Lanqing and Xu, Hang and Li, Zhenguo and Yeung, Dit-Yan and Kwok, James},
  journal={arXiv preprint arXiv:2310.05873},
  year={2023}
}

@inproceedings{gandikota2024unified,
  title={Unified concept editing in diffusion models},
  author={Gandikota, Rohit and Orgad, Hadas and Belinkov, Yonatan and Materzy{\'n}ska, Joanna and Bau, David},
  booktitle={Proceedings of the IEEE/CVF Winter Conference on Applications of Computer Vision},
  pages={5111--5120},
  year={2024}
}

@inproceedings{gong2024reliable,
  title={Reliable and efficient concept erasure of text-to-image diffusion models},
  author={Gong, Chao and Chen, Kai and Wei, Zhipeng and Chen, Jingjing and Jiang, Yu-Gang},
  booktitle={European Conference on Computer Vision},
  pages={73--88},
  year={2024},
  organization={Springer}
}

@article{tsai2023ring,
  title={Ring-a-bell! how reliable are concept removal methods for diffusion models?},
  author={Tsai, Yu-Lin and Hsu, Chia-Yi and Xie, Chulin and Lin, Chih-Hsun and Chen, Jia-You and Li, Bo and Chen, Pin-Yu and Yu, Chia-Mu and Huang, Chun-Ying},
  journal={arXiv preprint arXiv:2310.10012},
  year={2023}
}

@article{chin2023prompting4debugging,
  title={Prompting4debugging: Red-teaming text-to-image diffusion models by finding problematic prompts},
  author={Chin, Zhi-Yi and Jiang, Chieh-Ming and Huang, Ching-Chun and Chen, Pin-Yu and Chiu, Wei-Chen},
  journal={arXiv preprint arXiv:2309.06135},
  year={2023}
}

@inproceedings{zhang2024generate,
  title={To generate or not? safety-driven unlearned diffusion models are still easy to generate unsafe images... for now},
  author={Zhang, Yimeng and Jia, Jinghan and Chen, Xin and Chen, Aochuan and Zhang, Yihua and Liu, Jiancheng and Ding, Ke and Liu, Sijia},
  booktitle={European Conference on Computer Vision},
  pages={385--403},
  year={2024},
  organization={Springer}
}

@article{jain2024trasce,
  title={TraSCE: Trajectory Steering for Concept Erasure},
  author={Jain, Anubhav and Kobayashi, Yuya and Shibuya, Takashi and Takida, Yuhta and Memon, Nasir and Togelius, Julian and Mitsufuji, Yuki},
  journal={arXiv preprint arXiv:2412.07658},
  year={2024}
}

@article{zhang2024defensive,
  title={Defensive unlearning with adversarial training for robust concept erasure in diffusion models},
  author={Zhang, Yimeng and Chen, Xin and Jia, Jinghan and Zhang, Yihua and Fan, Chongyu and Liu, Jiancheng and Hong, Mingyi and Ding, Ke and Liu, Sijia},
  journal={Advances in Neural Information Processing Systems},
  volume={37},
  pages={36748--36776},
  year={2024}
}

@inproceedings{zhang2024forget,
  title={Forget-me-not: Learning to forget in text-to-image diffusion models},
  author={Zhang, Gong and Wang, Kai and Xu, Xingqian and Wang, Zhangyang and Shi, Humphrey},
  booktitle={Proceedings of the IEEE/CVF Conference on Computer Vision and Pattern Recognition},
  pages={1755--1764},
  year={2024}
}

@inproceedings{gandikota2023erasing,
  title={Erasing concepts from diffusion models},
  author={Gandikota, Rohit and Materzynska, Joanna and Fiotto-Kaufman, Jaden and Bau, David},
  booktitle={Proceedings of the IEEE/CVF International Conference on Computer Vision},
  pages={2426--2436},
  year={2023}
}

@article{brack2023sega,
  title={Sega: Instructing text-to-image models using semantic guidance},
  author={Brack, Manuel and Friedrich, Felix and Hintersdorf, Dominik and Struppek, Lukas and Schramowski, Patrick and Kersting, Kristian},
  journal={Advances in Neural Information Processing Systems},
  volume={36},
  pages={25365--25389},
  year={2023}
}

@inproceedings{kumari2023ablating,
  title={Ablating concepts in text-to-image diffusion models},
  author={Kumari, Nupur and Zhang, Bingliang and Wang, Sheng-Yu and Shechtman, Eli and Zhang, Richard and Zhu, Jun-Yan},
  booktitle={Proceedings of the IEEE/CVF International Conference on Computer Vision},
  pages={22691--22702},
  year={2023}
}

@article{kim2023towards,
  title={Towards safe self-distillation of internet-scale text-to-image diffusion models},
  author={Kim, Sanghyun and Jung, Seohyeon and Kim, Balhae and Choi, Moonseok and Shin, Jinwoo and Lee, Juho},
  journal={arXiv preprint arXiv:2307.05977},
  year={2023}
}

@article{sharma2024unlearning,
  title={Unlearning or concealment? a critical analysis and evaluation metrics for unlearning in diffusion models},
  author={Sharma, Aakash Sen and Sarkar, Niladri and Chundawat, Vikram and Mali, Ankur A and Mandal, Murari},
  journal={arXiv preprint arXiv:2409.05668},
  year={2024}
}

@article{zhu2024decoupling,
  title={Decoupling the Class Label and the Target Concept in Machine Unlearning},
  author={Zhu, Jianing and Han, Bo and Yao, Jiangchao and Xu, Jianliang and Niu, Gang and Sugiyama, Masashi},
  journal={arXiv preprint arXiv:2406.08288},
  year={2024}
}

@article{pham2024robust,
  title={Robust Concept Erasure Using Task Vectors},
  author={Pham, Minh and Marshall, Kelly O and Hegde, Chinmay and Cohen, Niv},
  journal={arXiv preprint arXiv:2404.03631},
  year={2024}
}

@article{sun2024attentive,
  title={Attentive Eraser: Unleashing Diffusion Model's Object Removal Potential via Self-Attention Redirection Guidance},
  author={Sun, Wenhao and Cui, Benlei and Tang, Jingqun and Dong, Xue-Mei},
  journal={arXiv preprint arXiv:2412.12974},
  year={2024}
}

@inproceedings{hao2025conceptexpress,
  title={ConceptExpress: Harnessing diffusion models for single-image unsupervised concept extraction},
  author={Hao, Shaozhe and Han, Kai and Lv, Zhengyao and Zhao, Shihao and Wong, Kwan-Yee K},
  booktitle={European Conference on Computer Vision},
  pages={215--233},
  year={2025},
  organization={Springer}
}

@article{gao2024meta,
  title={Meta-unlearning on diffusion models: Preventing relearning unlearned concepts},
  author={Gao, Hongcheng and Pang, Tianyu and Du, Chao and Hu, Taihang and Deng, Zhijie and Lin, Min},
  journal={arXiv preprint arXiv:2410.12777},
  year={2024}
}

@article{sridhar2024prompt,
  title={Prompt Sliders for Fine-Grained Control, Editing and Erasing of Concepts in Diffusion Models},
  author={Sridhar, Deepak and Vasconcelos, Nuno},
  journal={arXiv preprint arXiv:2409.16535},
  year={2024}
}

@article{maharanatowards,
  title={Towards Robust Concept Erasure in Diffusion Models: Unlearning Identity, Nudity and Artistic Styles},
  author={Maharana, Umakanta and Sharma, Aakash Sen and Sinha, Yash and Mali, Ankur and Kankanhalli, Mohan and Mandal, Murari}
}

@article{liu2024realera,
  title={RealEra: Semantic-level Concept Erasure via Neighbor-Concept Mining},
  author={Liu, Yufan and An, Jinyang and Zhang, Wanqian and Li, Ming and Wu, Dayan and Gu, Jingzi and Lin, Zheng and Wang, Weiping},
  journal={arXiv preprint arXiv:2410.09140},
  year={2024}
}

@article{han2024continuous,
  title={Continuous Concepts Removal in Text-to-image Diffusion Models},
  author={Han, Tingxu and Sun, Weisong and Hu, Yanrong and Fang, Chunrong and Zhang, Yonglong and Ma, Shiqing and Zheng, Tao and Chen, Zhenyu and Wang, Zhenting},
  journal={arXiv preprint arXiv:2412.00580},
  year={2024}
}

@article{yang2024pruning,
  title={Pruning for Robust Concept Erasing in Diffusion Models},
  author={Yang, Tianyun and Cao, Juan and Xu, Chang},
  journal={arXiv preprint arXiv:2405.16534},
  year={2024}
}

@inproceedings{pham2023circumventing,
  title={Circumventing concept erasure methods for text-to-image generative models},
  author={Pham, Minh and Marshall, Kelly O and Cohen, Niv and Mittal, Govind and Hegde, Chinmay},
  booktitle={The Twelfth International Conference on Learning Representations},
  year={2023}
}

@article{heusel2017gans,
  title={Gans trained by a two time-scale update rule converge to a local nash equilibrium},
  author={Heusel, Martin and Ramsauer, Hubert and Unterthiner, Thomas and Nessler, Bernhard and Hochreiter, Sepp},
  journal=NIPS,
  volume={30},
  year={2017}
}

@inproceedings{zhang2018unreasonable,
  title={The unreasonable effectiveness of deep features as a perceptual metric},
  author={Zhang, Richard and Isola, Phillip and Efros, Alexei A and Shechtman, Eli and Wang, Oliver},
  booktitle={Proceedings of the IEEE conference on computer vision and pattern recognition},
  pages={586--595},
  year={2018}
}

@inproceedings{he2016deep,
  title={Deep residual learning for image recognition},
  author={He, Kaiming and Zhang, Xiangyu and Ren, Shaoqing and Sun, Jian},
  booktitle={Proceedings of the IEEE conference on computer vision and pattern recognition},
  pages={770--778},
  year={2016}
}

@article{howard2020fastai,
  title={Fastai: A layered API for deep learning},
  author={Howard, Jeremy and Gugger, Sylvain},
  journal={Information},
  volume={11},
  number={2},
  pages={108},
  year={2020},
  publisher={Multidisciplinary Digital Publishing Institute}
}

@misc{bedapudi2019nudenet,
  title={Nudenet: Neural nets for nudity classification, detection and selective censoring},
  author={Bedapudi, P},
  year={2019}
}

@article{xing2024survey,
  title={A survey on video diffusion models},
  author={Xing, Zhen and Feng, Qijun and Chen, Haoran and Dai, Qi and Hu, Han and Xu, Hang and Wu, Zuxuan and Jiang, Yu-Gang},
  journal={ACM Computing Surveys},
  volume={57},
  number={2},
  pages={1--42},
  year={2024},
  publisher={ACM New York, NY}
}

@inproceedings{li2023q,
  title={Q-diffusion: Quantizing diffusion models},
  author={Li, Xiuyu and Liu, Yijiang and Lian, Long and Yang, Huanrui and Dong, Zhen and Kang, Daniel and Zhang, Shanghang and Keutzer, Kurt},
  booktitle={Proceedings of the IEEE/CVF International Conference on Computer Vision},
  pages={17535--17545},
  year={2023}
}

@inproceedings{zhang2023adding,
  title={Adding conditional control to text-to-image diffusion models},
  author={Zhang, Lvmin and Rao, Anyi and Agrawala, Maneesh},
  booktitle={Proceedings of the IEEE/CVF International Conference on Computer Vision},
  pages={3836--3847},
  year={2023}
}

@inproceedings{ma2024deepcache,
  title={Deepcache: Accelerating diffusion models for free},
  author={Ma, Xinyin and Fang, Gongfan and Wang, Xinchao},
  booktitle={Proceedings of the IEEE/CVF Conference on Computer Vision and Pattern Recognition},
  pages={15762--15772},
  year={2024}
}

@inproceedings{somepalli2023diffusion,
  title={Diffusion art or digital forgery? investigating data replication in diffusion models},
  author={Somepalli, Gowthami and Singla, Vasu and Goldblum, Micah and Geiping, Jonas and Goldstein, Tom},
  booktitle={Proceedings of the IEEE/CVF Conference on Computer Vision and Pattern Recognition},
  pages={6048--6058},
  year={2023}
}

@inproceedings{corvi2023intriguing,
  title={Intriguing properties of synthetic images: from generative adversarial networks to diffusion models},
  author={Corvi, Riccardo and Cozzolino, Davide and Poggi, Giovanni and Nagano, Koki and Verdoliva, Luisa},
  booktitle={Proceedings of the IEEE/CVF Conference on Computer Vision and Pattern Recognition},
  pages={973--982},
  year={2023}
}

@inproceedings{zhang2024artbank,
  title={ArtBank: Artistic Style Transfer with Pre-trained Diffusion Model and Implicit Style Prompt Bank},
  author={Zhang, Zhanjie and Zhang, Quanwei and Xing, Wei and Li, Guangyuan and Zhao, Lei and Sun, Jiakai and Lan, Zehua and Luan, Junsheng and Huang, Yiling and Lin, Huaizhong},
  booktitle={Proceedings of the AAAI Conference on Artificial Intelligence},
  volume={38},
  number={7},
  pages={7396--7404},
  year={2024}
}

@inproceedings{liu2024structure,
  title={Structure Matters: Tackling the Semantic Discrepancy in Diffusion Models for Image Inpainting},
  author={Liu, Haipeng and Wang, Yang and Qian, Biao and Wang, Meng and Rui, Yong},
  booktitle={Proceedings of the IEEE/CVF Conference on Computer Vision and Pattern Recognition},
  pages={8038--8047},
  year={2024}
}

@article{yue2024resshift,
  title={Resshift: Efficient diffusion model for image super-resolution by residual shifting},
  author={Yue, Zongsheng and Wang, Jianyi and Loy, Chen Change},
  journal={Advances in Neural Information Processing Systems},
  volume={36},
  year={2024}
}

@inproceedings{fan2025adadiffsr,
  title={Adadiffsr: Adaptive region-aware dynamic acceleration diffusion model for real-world image super-resolution},
  author={Fan, Yuanting and Liu, Chengxu and Yin, Nengzhong and Gao, Changlong and Qian, Xueming},
  booktitle={European Conference on Computer Vision},
  pages={396--413},
  year={2025},
  organization={Springer}
}

@inproceedings{li2024privimage,
  title={$\{$PrivImage$\}$: Differentially Private Synthetic Image Generation using Diffusion Models with $\{$Semantic-Aware$\}$ Pretraining},
  author={Li, Kecen and Gong, Chen and Li, Zhixiang and Zhao, Yuzhong and Hou, Xinwen and Wang, Tianhao},
  booktitle={33rd USENIX Security Symposium (USENIX Security 24)},
  pages={4837--4854},
  year={2024}
}

@inproceedings{yang2024mma,
  title={Mma-diffusion: Multimodal attack on diffusion models},
  author={Yang, Yijun and Gao, Ruiyuan and Wang, Xiaosen and Ho, Tsung-Yi and Xu, Nan and Xu, Qiang},
  booktitle={Proceedings of the IEEE/CVF Conference on Computer Vision and Pattern Recognition},
  pages={7737--7746},
  year={2024}
}

@inproceedings{zhu2024watermark,
  title={Watermark-embedded Adversarial Examples for Copyright Protection against Diffusion Models},
  author={Zhu, Peifei and Takahashi, Tsubasa and Kataoka, Hirokatsu},
  booktitle={Proceedings of the IEEE/CVF Conference on Computer Vision and Pattern Recognition},
  pages={24420--24430},
  year={2024}
}

@article{yangfederated,
  title={Federated Unlearning with Diffusion Models},
  author={Yang, Mingzhao and Li, Bin and Xue, Xiangyang}
}

@article{lyu2023one,
  title={One-dimensional Adapter to Rule Them All: Concepts, Diffusion Models and Erasing Applications},
  author={Lyu, Mengyao and Yang, Yuhong and Hong, Haiwen and Chen, Hui and Jin, Xuan and He, Yuan and Xue, Hui and Han, Jungong and Ding, Guiguang},
  journal={arXiv preprint arXiv:2312.16145},
  year={2023}
}

\end{document}